\documentclass[12pt]{iopart}

\newcommand{\eqref}[1]{(\ref{#1})}

\usepackage{iopams}
\usepackage{graphicx}
\usepackage{amsfonts}
\usepackage{amsbsy}
\usepackage{amssymb}

\usepackage[top=2cm,bottom=2cm,left=1.5cm,right=1.5cm,columnsep=0.7cm]{geometry}

\usepackage[T1]{fontenc}
\usepackage[utf8]{inputenc}
\usepackage{enumerate}   
\usepackage{bm}

\usepackage{microtype}

\usepackage{color}

\usepackage{array}

\usepackage{citesort}

\begin{document}

\title{The Entropy Production of Ornstein-Uhlenbeck Active Particles: \\a path integral method for correlations}

\author{Lorenzo Caprini$^1$, Umberto Marini Bettolo Marconi$^2$, Andrea Puglisi$^3$, Angelo Vulpiani$^{4,5}$ }



\address{$^1$ Gran Sasso Science Institute (GSSI), Via. F. Crispi 7, 67100 L'Aquila, Italy}
\address{$^2$ Scuola di Scienze e Tecnologie, Universit\`a di Camerino - via Madonna delle Carceri, 62032, Camerino, Italy}
\address{$^3$ CNR-ISC, Consiglio Nazionale delle Ricerche}
\address{$^4$ Dipartimento di Fisica, Universit\`a Sapienza - p.le A. Moro 2, 00185, Roma, Italy}
\address{$^5$ Centro Interdisciplinare "B.Segre", Accademia dei Lincei, Roma, Italy}


\ead{ lorenzo.caprini@gssi.it}

\vspace{10pt}
\begin{indented}
\item[]Gennuary 2019
\end{indented}

\begin{abstract}
By employing a path integral formulation, we obtain the
entropy production rate for a system of active Ornstein-Uhlenbeck
particles (AOUP) both in the presence and in the absence of thermal
noise.  The present treatment clarifies some contraddictions
concerning the definition of the entropy production rate in the AOUP
model, recently appeared in the literature. We derive explicit formulas for three different cases:
overdamped Brownian particle, AOUP with and without thermal noise. 
In addition, we show that it is not necessary to introduce additional hypotheses
concerning the parity of auxiliary variables under time reversal
transformation. Our results agree with those based on a previous
mesoscopic approach.
\end{abstract}


\maketitle

\section{Introduction}
 In recent years, the rapid development of nanotechnologies opened the possibility of manipulating matter at the single molecule level and drew the attention of the physicists towards the study of systems comprising few particles \cite{rogers2014nanotechnology}. Due to the smallness of the systems considered, the standard thermodynamic description is insufficient  and it is necessary to take into account fluctuations around the average values of thermodynamic variables. 
Stochastic energetics \cite{sekimoto2010stochastic,seifert2012stochastic,sekimoto1998langevin,jarzynski2011equalities} has been introduced to deal with these requirements: it extends thermodynamic concepts to single particle trajectories and generalizes familiar quantities such as internal energy, exchanged heat, work and entropy to systems 
consisting of few particles or even a single particle.
Such an approach introduces the concept of trajectory as the time history of the system and associates to it the notion of
entropy production of a single realization of the dynamics \cite{gallavotti1995dynamical,lebowitz1999gallavotti}. An average over different realizations reproduces the macroscopic values of the thermodynamic variables. 
The physical interpretation of this methodology has been discussed in \cite{seifert2005entropy,speck2007distribution,seifert2012stochastic}, where the concepts of entropy production of the medium, i.e the entropy flux to the environment, and the total entropy production of the system have been connected.
In particular, the entropy production has been identified as the energy dissipated in the thermostat, whose form generally relies on identifying thermal forces from the dynamics.

Some years ago, the theory has been generalized to include non-Markovian processes
characterized by memory effects in the noise and/or in the memory kernel \cite{zamponi2005fluctuation}. 
More recently, more complicated cases have been studied where it has been proposed 
to modify  somehow ``arbitrarily"  the definition of entropy production rate~\cite{munakata1,munakata2,kim,ganguly2013stochastic,mandal2018}. In certain cases, this is a meaningful procedure, for instance  in the presence of external magnetic fields \cite{pradhan2010nonexistence,andrieux2008quantum,chun2018microscopic,kwon2016unconventional}, where employing modified backward-trajectory generators leads to expected and well-known results. In other cases the meaning of these modified entropy production definitions is less clear, see for instance the discussion in~\cite{cerino}.
A recent example where such a trajectory approach to the entropy production has been  employed 
and generated some debate is the case of active systems~\cite{ganguly2013stochastic,nardini2017entropy,marconi2017heat,puglisi2017clausius,mandal2018,caprini2018comment,dabelow2018irreversibility,shankar2018hidden}.
These systems include a large class of self-propelled agents, which live at different length scales
and share a peculiar feature, namely the ability to convert energy from the environment into directed persistent motion.
In particular, a single active agent on a small time scale shows a persistent trajectory, caused by some chemical reactions
in the case of a manmade artificial Janus particle \cite{Lattuada2011Synthesis,walther2013janus}
 or by biological complex mechanisms such as flagella in the case of natural Escherichia coli \cite{berg2008coli}. These mechanisms consume 
energy and render active systems intrinsically out of equilibrium. Since we expect a non-zero entropy production in their steady state even in absence of external forces
it is quite natural to explore their dynamics under the lens of stochastic thermodynamics with the idea of 
gaining better control of their functioning and finding strategies to optimize their performances~\cite{ramaswamy2010mechanics,bechinger2016active,marchetti2013hydrodynamics,romanczuk2012active}.

Several descriptions have been developed in order to capture the behavior of active systems: among the simplest models we  mention the Run and Tumble \cite{nash2010run,tailleur2008statistical,angelani2015run,sevilla2018stationary} and the Active Brownian Particles (ABP) model \cite{ten2011brownian,romanczuk2012active,sevilla2015smoluchowski}. More recently, the active Ornstein-Uhlenbeck particle (AOUP) \cite{szamel2014self,maggi2015multidimensional} model has been introduced as a convenient approximation of the ABP\cite{caprini2018activity,das2018confined}. In fact, the former leads to a simpler analytical treatment and
gives more possibilities to obtain interesting predictions \cite{farage2015effective,marconi2016velocity,caprini2018linear,caprini2018activeescape}.

The computation of the entropy production in the AOUP model has originated some debate concerning the following question: what is the parity of the self-propelling force under time-reversal transformation (TRT)? 
 Assuming that self-propulsion is even or odd, several authors obtain different results for the entropy production, as addressed in \cite{dabelow2018irreversibility}.
In the present work, we show that a hypothesis about the time-reversal parity of the self-propelling force is not necessary \cite{caprini2018comment} and propose a different method to compute the  entropy production, which basically coincides with the non-Markovian generalization of the path integral approach~\cite{zamponi2005fluctuation}. 

Moreover, we discuss the  more realistic case where, in addition to the stochastic active force,
 another source of noise, that is the thermal bath modelling the solvent medium, is included in the dynamics.
This possibility has been already explored considering both ABP \cite{shankar2018hidden,bandopadhyay2017rotational,speck2016stochastic,chaudhuri2014active} and AOUP dynamics \cite{puglisi2017clausius}, always taking some assumption about the parity of the self-propulsion force. Again we show here that a univocal computation is possible.

We mention that other models have been recently considered, for instance taking explicitly into account the role of hydrodynamic interactions \cite{seifert2018stochastic}, chirality \cite{pigolotti2017generic} or by considering explicitly the chemical nature of the self-propulsion force\cite{dadhichi2018origins,speck2018active}.
Entropy production has been recently computed using Field Theories models without Time-Reversal Symmetry \cite{nardini2017entropy}.

The article is structured as follows: in Sec.~\ref{pathintegral} we briefly review the general path-integral approach in the presence of a non-Markovian noise, discussing a generalization of Sekimoto's formula for the injected power. In Section~\ref{sec:overdampedExamples} we provide some examples: for completeness, we apply the method to the well-known result for a Brownian colloid in a thermal bath and discuss the case of the AOUP dynamics with and without the presence of the thermal noise, i.e. considering or not the role of the thermal environment. 
Finally, in Sec.~\ref{sec:conclusion}, we summarize the results and present some conclusions.

\section{Path Integral approach}
\label{pathintegral}
To introduce the path integral approach, we consider the more general dynamics in terms of the set of state variables, $\boldsymbol{\omega}$, describing our physical system.
Without any loss of generality, we can express the dynamics as a set of first order differential equations, whose particular structure depends on the physical properties of our model: 
\begin{equation}
\label{eq:generaldynamics}
\dot{\boldsymbol{\omega}}=\boldsymbol{F}(\boldsymbol{\omega}) + \boldsymbol{\eta} \,.
\end{equation}
The vector force,
$\boldsymbol{F}$, is a generic vector function, depending in principle on all the variables
 and containing the deterministic contribution to the dynamics.  Instead, the vector $\boldsymbol{\eta}$ contains the stochastic 
 contributions to the evolution and  could be a signal correlated in time. In the picture of mesoscopic dynamics, we can imagine that $\boldsymbol{\eta}$, hereafter referred as noise vector, stems from a coarse-grained procedure of some degrees of freedom whose dynamics is faster  than the one associated with the slow variables, $\boldsymbol{\omega}$.
To fix the ideas, for an underdamped colloidal particle in a thermal bath the set $\boldsymbol{\omega}$ can be identified with the position, $\boldsymbol{x}$, and the velocity, $\boldsymbol{v}$, of  the particle.  
As stated in the Introduction, there exist many physical systems where the noise is correlated in time and/or displays some correlations between its components.
For the sake of simplicity, we assume that the first noise moment, $\langle \boldsymbol{\eta} \rangle$, is zero, without any loss of generality, and the second moment is given by the two-time correlation matrix, 
\begin{equation*}
\nu_{ij}(t-s)=\langle \eta_i(s)\eta_j(t)\rangle.
\end{equation*}
For instance, in the so-called AOUP model exponentially decaying correlations appear in the noise. In order to exhibit the generality of such a result, we consider a general invertible matrix $\boldsymbol{\nu}(t-s)$ in this Section.

In the following, we use the compact notation, $\underline{\boldsymbol{\omega}}=\{\boldsymbol{\omega}\}_{t_0}^{\mathcal{T}}$, to denote the time history of the single trajectory between the initial time, $t_0$, and the final time, $\mathcal{T}$.
The explict introduction of a source of noise in the  dynamics, produces a non-trivial probability, $P[\underline{\boldsymbol{\omega}}|\boldsymbol{\omega}_0 ]$, of observing a path $\underline{\boldsymbol{\omega}}$ given the intial state $\boldsymbol{\omega}_0$.
In the following, we consider Gaussian noises, which are entirely specified by the mean values and correlations, $\langle \boldsymbol{\eta}\rangle$ and $\nu_{ij}(t-s)$. 
Under these assumptions, the probability of observing the noise path, $\underline{\boldsymbol{\eta}}$, reads:
\begin{equation}
\label{eq:noiseprob}
\tilde{P}[\underline{\boldsymbol{\eta}}|\boldsymbol{\eta}_0] \propto\exp{ \left[-\frac{1}{2}\int dt \int ds \,\boldsymbol{\eta}(s)\boldsymbol{T}^{-1}(t-s)\boldsymbol{\eta}(t) \right] } ,
\end{equation}
where we dropped an irrelevant normalization factor.  
The two-time integrals of Eq.~\eqref{eq:noiseprob} 
are computed along the times involved in the evolution of the trajectory, from $t_0$ to $\mathcal{T}$.
 The operator $\boldsymbol{T^{-1}}$ is the inverse of the two time-correlator $\boldsymbol{\nu}$, defined as 
\begin{equation*}
\int dt' \boldsymbol{T}^{-1}(t-t') \boldsymbol{\nu}(t'-s) =\boldsymbol{I} \delta(t-s),
\end{equation*}
where $\boldsymbol{I}$ is the unit matrix and $\delta(t)$ is the Dirac delta function.
In the Appendix \ref{appendixa}, we compute  $\boldsymbol{T}^{-1}$ in Fourier space for some choices of noise correlations.

Using Eq.~\eqref{eq:generaldynamics} we express the noise in terms of $\boldsymbol{\omega}$ and $\dot{\boldsymbol{\omega}}$ through a change of variables.  Formally, we have $\boldsymbol{\eta}=\boldsymbol{ \eta}[\boldsymbol{\omega},\dot{\boldsymbol{\omega}}]$ from Eq.\eqref{eq:generaldynamics} and the following relation between the two path probabilities:
\begin{equation}
\log{P[\underline{\boldsymbol{\omega}}|\boldsymbol{\omega}_0 ]}= \log{\tilde{P}[\underline{\boldsymbol{\eta}}|\boldsymbol{\eta}_0]} +\log{\det{\frac{\partial \underline{\boldsymbol{\eta}}}{\partial \underline{\boldsymbol{\omega}}}}} \,,
\end{equation}
where the last term is the Jacobian of the transformation from $\underline{\boldsymbol{\eta}}$  to $\underline{\boldsymbol{\omega}}$.
Therefore, we can express the probability of a trajectory in terms of both the dynamical variables of the system and their time variations:
\begin{equation}
\label{eq:directprob}
P[\underline{\boldsymbol{\omega}}|\boldsymbol{\omega}_0 ]\propto \exp{\left[-\frac{1}{2}\int dt\int ds \,\,\boldsymbol{ \eta}[\boldsymbol{\omega}(s),\dot{\boldsymbol{\omega}}(s)] \boldsymbol{T}^{-1}(t-s) \,\,\boldsymbol{ \eta}[\boldsymbol{\omega}(t),\dot{\boldsymbol{\omega}}(t)]\right]} \,.
\end{equation}
In this path probability, we have neglected the contribution of the determinant, since 
under some rather general conditions \cite{spinney2012nonequilibrium,spinney2012entropy,chaudhuri2016entropy} it does not play any role in the computation of the entropy production of the system. 
By taking into account the relation \eqref{eq:generaldynamics}, without loss of generality, Eq.\eqref{eq:directprob} can be rewritten as:
\begin{equation}
P[\underline{\boldsymbol{\omega}}|\boldsymbol{\omega}_0 ]\propto \exp{\left[-\frac{1}{2}\int dt \int ds \left[\dot{\boldsymbol{\omega}} - \boldsymbol{F}(\boldsymbol{\omega})\right](s)
 \boldsymbol{T}^{-1}(t-s) \,\, \left[\dot{\boldsymbol{\omega}} - \boldsymbol{F}(\boldsymbol{\omega})\right](t)\right]} \,.
\end{equation}


\subsection{Reversed trajectory}
In order to obtain the entropy production rate, we have to determine the probability associated with 
the reversed trajectory. To do so, we need  to know how each variable transforms under a time reversal transformation. 
In the following, we denote with $\boldsymbol{\Theta}$ the time-reversal operator, which acts on a time-dependent observable $o(t)$ as $\boldsymbol{\Theta} o(t) = o(\mathcal{T}-t)$. We remark that in our treatment
each component of $\boldsymbol{\omega}$ does not have to possess necessarily a definite parity under time reversal (odd or even), but can
transform as a combination of odd or even variables according to some fixed prescription.

Restricting to even or odd variables, the probability associated with the reversed path is:
\begin{equation}
   P\left[\boldsymbol{\Theta}  \underline{\boldsymbol{\omega}} |\boldsymbol{\Theta}\boldsymbol{\omega}_0 \right]   
\propto \exp{\left[-\frac{1}{2}\int dt \int ds \left[-\boldsymbol{\epsilon}\dot{\boldsymbol{\omega}} - \boldsymbol{F}(\boldsymbol{\epsilon}\boldsymbol{\omega})\right](s)
 \boldsymbol{T}^{-1}(t-s) \,\, \left[-\boldsymbol{\epsilon}\dot{\boldsymbol{\omega}} - \boldsymbol{F}(\boldsymbol{\epsilon}\boldsymbol{\omega})\right](t)  \right]},
\end{equation}
where $\boldsymbol{\epsilon}$ is a diagonal matrix with elements $\pm1$ for even and odd components under time reversal symmetry, respectively, and
 $\boldsymbol{\Theta}\boldsymbol{\omega}_0=\boldsymbol{\epsilon}\boldsymbol{\omega}_{\mathcal{T}}$ .

 In the following, we assume that $\boldsymbol{\omega}$ contains only even variables under time reversal
transformation (TRT), in such a way that $\boldsymbol{\epsilon}$ reduces to the identity matrix. 
Since by definition $\boldsymbol{\nu}$ is even under TRT, it follows that also $\boldsymbol{T}^{-1}$ is even.
In this case, the  probability of the reversed trajectory reads:
\begin{equation}
P\left[\boldsymbol{\Theta}  \underline{\boldsymbol{\omega}} |\boldsymbol{\Theta}\boldsymbol{\omega}_0 \right]   
\propto \exp{\left[-\frac{1}{2}\int dt \int ds \left[-\dot{\boldsymbol{\omega}} - \boldsymbol{F}(\boldsymbol{\omega})\right](s)
 \boldsymbol{T}^{-1}(t-s) \,\, \left[-\dot{\boldsymbol{\omega}} - \boldsymbol{F}(\boldsymbol{\omega})\right](t)\right]} \,.
\end{equation}

\subsection{Entropy Production and Dissipation}
After these preliminaries,
following~\cite{zamponi2005fluctuation} we are ready to compute the entropy production of the medium, $\Sigma_{\mathcal{T}}$, in terms of the ratio between the probability of the forward and backward trajectory.
\begin{eqnarray}
\label{eq:generalentropyprod}
\Sigma_{\mathcal{T}} &&=\log{\frac{P[\underline{\boldsymbol{\omega}}|\boldsymbol{\omega}_0]}{P{[\boldsymbol{\Theta}{\underline{\boldsymbol{\omega}}|\boldsymbol{\Theta} \boldsymbol{\omega}_0]}}}}= \frac{1}{2} \int dt \int ds \,\, \biggl[\dot{\boldsymbol{\omega}}(t)\boldsymbol{T}^{-1}(t-s)\boldsymbol{F}(\boldsymbol{\omega}(s))  + \boldsymbol{F}(\boldsymbol{\omega}(t))\boldsymbol{T}^{-1}(t-s)\dot {\boldsymbol{\omega} }(s)      \biggr] \nonumber\\
&& + \frac{1}{2}\int dt \int ds \,\, \biggl[\boldsymbol{\epsilon}\,\dot{\boldsymbol{\omega}}(t)\boldsymbol{T}^{-1}(t-s)\boldsymbol{F}(\boldsymbol{\epsilon}\,\boldsymbol{\omega}(s))  + \boldsymbol{F}(\boldsymbol{\epsilon}\,\boldsymbol{\omega}(t))\boldsymbol{T}^{-1}(t-s)\,\boldsymbol{\epsilon}\,\dot{\boldsymbol{\omega}}(s)      \biggr] \,.
\end{eqnarray}
In the first passage we have dropped the quadratic terms $\boldsymbol{\omega} \boldsymbol{T}^{-1}\boldsymbol{\omega}$ and $\boldsymbol{F} \boldsymbol{T}^{-1} \boldsymbol{F}$ since they trivially contribute as boundary terms, as shown in \cite{zamponi2005fluctuation}. 
In the case of even variables  $\boldsymbol{\omega}$,  the two terms in the right-hand side of Eq.~\eqref{eq:generalentropyprod} are equal and rewrite:
\begin{eqnarray}
\label{eq:generalentropyprodeven}
\Sigma_{\mathcal{T}} &&=
\log{\frac{P{[\underline{\boldsymbol{\omega}}|\boldsymbol{\omega}_0]}}{P{[\boldsymbol{\Theta}{\underline{\boldsymbol{\omega}}|\boldsymbol{\Theta} \boldsymbol{\omega}_0]}}}}=  \int dt \int ds \,\, \biggl[\dot{\boldsymbol{\omega}}(t)\boldsymbol{T}^{-1}(t-s)\boldsymbol{F}(\boldsymbol{\omega}(s))  + \boldsymbol{F}(\boldsymbol{\omega}(t))\boldsymbol{T}^{-1}(t-s)\dot{\boldsymbol{\omega}}(s)      \biggr]\nonumber\\
&&=  \int dt \,\,\biggl[\dot{\boldsymbol{\omega}}(t) \left(\boldsymbol{T}^{-1} \ast \boldsymbol{F}\right)(t)  + \boldsymbol{F}(t) \left(\boldsymbol{T}^{-1} \ast \dot{\boldsymbol{\omega}}\right)(t) \biggr] \,,
\end{eqnarray}
where the symbol $\ast$ denotes the convolution operation. Indeed, since $T^{-1}$ is a decreasing function of its argument, we can extend one of the integrals between $-\infty$ to $\infty$, just by producing a subdominant term which disappears in the steady states obtained for large $\mathcal{T}$.
This allows us to define the time-dependent entropy production rate as:
\begin{equation}
\label{eq:entropyprodrate}
\sigma(t) =  \dot{\boldsymbol{\omega}}(t) \left(\boldsymbol{T}^{-1} \ast \boldsymbol{F} \right)(t)+\boldsymbol{F}(t) \left(\boldsymbol{T}^{-1} \ast \dot{\boldsymbol{\omega}}   \right)(t)\,.
\end{equation}
The first term of Eq.\eqref{eq:entropyprodrate}, apart from a factor $1/2$, has the same form of the generalized Sekimotos' injection 
term, while the second addend is new. In the Brownian case  since $\boldsymbol{T}^{-1}(t-s)\propto\boldsymbol{I} \delta(t-s)$ the two terms are equal, but
in general, they are not. 
If  $\boldsymbol{T}^{-1}(t-s)=\delta(t-s)\boldsymbol{G}^{-1}(t)$ is an operator local in time, Eq.\eqref{eq:entropyprodrate} reduces to
the following form:
\begin{equation}
\label{eq:general_injectedpower}
\sigma(t) = \biggl[ \dot{\boldsymbol{\omega}} \,\boldsymbol{G}^{-1} \,\boldsymbol{F}+\boldsymbol{F}\, \boldsymbol{G}^{-1}  \,\dot{\boldsymbol{\omega}} \biggr](t) = \biggl[\dot{\boldsymbol{\omega}} \, \boldsymbol{G}^{-1}  \, \boldsymbol{F} + \left(\dot{\boldsymbol{\omega}} \, \boldsymbol{G}^{-1}  \, \boldsymbol{F}\right)^{adj} \biggr] (t) \,,
\end{equation}
where  the superscript $adj$ stands for the adjoint operation and in the last equality we have used that $\boldsymbol{T}^{-1}=\left(\boldsymbol{T}^{-1}\right)^{adj}$. We point out that even when  $\boldsymbol{T}^{-1}(t-s)=\delta(t-s)\boldsymbol{G}^{-1}(t)$ , still
 $\boldsymbol{T}^{-1}$ can represent a differential and not a multiplicative operator.

 It is interesting to connect the entropy production to the dissipation, $I$, defined as the imbalance between the power dissipated  by the drag force and the power injected by the noise. 
This definition, originally developed by Sekimoto for driven Langevin processes \cite{sekimoto1997complementarity,sekimoto1997kinetic} have been generalized to particles immersed into a viscoelastic bath \cite{Fodor2016EplNoneqDiss}.
In this framework, the generalization of the dissipation to systems following the dynamics \eqref{eq:generaldynamics} is straightforward: 
 \begin{equation}
\label{eq:simplified_InjectedPower}
 I= \dot{\boldsymbol{\omega}} \, \boldsymbol{G}^{-1}  \, \boldsymbol{F} =- \dot{\boldsymbol{\omega}} \, \boldsymbol{G}^{-1}\,\dot{\boldsymbol{\omega}} + \dot{\boldsymbol{\omega}} \, \boldsymbol{G}^{-1}\, \eta \,.
\end{equation}
 We remark that the form of $I$ resembles the dissipation of particles immersed in a viscoelastic bath.
 Now, we can express the entropy production rate in terms of the dissipation:
 \begin{equation}
\label{eq:general_inj}
\sigma
=   I+ I^{adj}  \,.
\end{equation}
Eq.\eqref{eq:general_inj} generalizes  Sekimoto's result  for a Brownian dynamics where $I=I^{adj}$, which will be explicitly reviewed in the next Section.

Finally, assuming the ergodicity
the time average of $\sigma(t) $ in the limit ${\mathcal{T}}\to \infty $ is the same as its average over the probability space
thus we can write Eq.~\eqref{eq:generalentropyprodeven} as:
\begin{equation}
\label{eq:generalentropyprodeven2}
\Sigma_{\mathcal{T}} \simeq {\mathcal{T}} \,\langle \sigma\rangle = {\mathcal{T}}\,\int d\boldsymbol{\omega} p_s({\boldsymbol{\omega}}) \,\,\biggl[\dot{\boldsymbol{\omega}} \left(\boldsymbol{T}^{-1} \ast \boldsymbol{F}\right)  + \boldsymbol{F} \left(\boldsymbol{T}^{-1} \ast \dot{\boldsymbol{\omega}}\right) \biggr] \,.
\end{equation}
where $\simeq$ means that we are considering the entropy production in the stationary state. In other words, we do not include the time-dependent entropy production due to typical time to reach the steady state, which depends on the initial conditions.

\section{Over-damped dynamics}\label{sec:overdampedExamples}
In this Section, we study the entropy production of particles in the over-damped regime and neglect the inertial terms. 
 For the sake of simplicity, we restrict to the one-dimensional case.
According to such a choice,
the dynamics is described by the set $\boldsymbol{\omega}$ containing only the particle position, $\boldsymbol{x}$, which are even under time-reversal transformation. In particular, the dynamics takes the simple form:
\begin{equation}
\label{eq:generaloverdamped}
\dot{x} = \frac{F}{\gamma} + \eta, \qquad F = - \Psi'(x) \,,
\end{equation}
where $F$ is the force due to the external potential, $\Psi(x)$, and $\gamma$ the drag coefficient. 
We can easily extend the theory to the case of a driving deterministic force which intrinsically pushes the particle out of equilibrium, a generalization which does not add anything to the current discussion.
The stochastic source, $\eta$, can represent the contribution of the solvent surrounding the tagged colloidal particle, as usual for Brownian dynamics, but also the internal self-propulsion mechanism of microswimmers, as recently assumed in active matter models. 
Other interpretations of such a model, proposed by Di Leonardo et. al.\cite{maggi2014generalized,maggi2017memory} and experimentally confirmed, suggest to consider the noise amplitude as the effect of a bath of active particles on a tracer passive body, whose evolution is described by Eq.\eqref{eq:generaloverdamped}. In particular, choosing $F$ as a force due to an external harmonic potential we can model the effect of an active gel \cite{fodor2014energetics,ben2015modeling,fodor2015activity,chaki2018entropy}, as observed experimentally \cite{e2014time}.

\subsection{Example I: Brownian particle}
Let us begin with the case of a Brownian particle in the presence of a confining potential.
Its evolution is described by the over-damped dynamics \eqref{eq:generaloverdamped}, where $\eta$ is a Gaussian noise vector with zero-mean and variance $\langle\eta(t)\eta(s) \rangle=2(T_b/\gamma)\delta(t-s)$. 
According to this choice, we have:
\begin{equation}
T^{-1}(t) =\gamma \frac{\delta(t)}{2T_b} 
\end{equation}
and the entropy production rate, using  Eq.~\eqref{eq:entropyprodrate}, is:
\begin{equation}
\label{sigmabrown}
\sigma(t) = -\frac{\gamma}{2T_b}\left[ \frac{ \Psi'}{\gamma} \dot{x}  +  \dot{x} \frac{ \Psi'}{\gamma}  \right] = -\frac{\dot{x}}{ T_b} \Psi'  =-\frac{1}{ T_b} \frac{d}{dt}\Psi .
\end{equation}
The last equality shows that $\sigma(t)$ is a time derivative and this
represents a boundary term (b.t.) with the consequence that the average entropy production \eqref{eq:generalentropyprodeven} vanishes:
\begin{equation}
\Sigma_{\mathcal{T}} \simeq {\mathcal{T}} \,\langle \sigma\rangle  = -\int \frac{1}{ T_b} \frac{d}{dt}\Psi  \,dt= b.t.
\end{equation}
This result is consistent with the validity of the detailed balance condition: in fact, the particle is globally in equilibrium with the environment and there is no production of entropy.

Moreover, according to Eq.\eqref{eq:simplified_InjectedPower} the injected power, $I$, reads:
\begin{equation}
\label{eq:BrownianInjectedpower}
I= \dot{x}\,  \Psi' = \gamma \dot{x}\left(\dot{x} -  \eta\right) \,,
\end{equation}
which is nothing but Sekimoto's result and provides a consistency check for our approach.  Taking the stationary average of Eq.\eqref{eq:BrownianInjectedpower}, it is straightforward to see that $\langle I \rangle= \frac{d}{dt} \langle \Psi\rangle$ which necessarily vanishes in the steady-state, as we expect.

\subsection{Example II: Self-propelled particles without thermal noise}
\label{notranslational}
Let us turn to study an example of far from equilibrium system, inspired by the physics of active matter.
Due to the nature of the self-propulsion, we expect to encounter a non-zero entropy production.

We shall employ the AOUP dynamics in the absence of a thermal bath, i.e. in a regime where the thermal noise due to the environment in which the microswimmers are immersed is not considered. Such an assumption is based on experimental observations  of several systems, for which the diffusion due to the thermal noise is negligible with respect to the one due to the self-propulsion. 
 The equation of motion for the position, $\mathbf{x}$,  is still given by the overdamped equation  Eq.\eqref{eq:generaloverdamped} but in order to capture the physics of complex microswimmers such as E.Coli \cite{berg2008coli}, protozoa \cite{blake1974mechanics} or living tissues \cite{poujade2007collective}
 the model contains an extra degree of freedom. We refer to this degree of freedom as self-propulsion force or simply self-propulsion. In the AOUP, the self-propulsion is obtained by replacing the white-noise source $\eta$ in Eq.\eqref{eq:generaloverdamped} 
 by a Gaussian noise, $\eta_a$, exponentially correlated in time. For consistency with our previous works, 
 we choose the two-time correlation of $\eta_a$ to be
\begin{equation*}
\langle\eta_a(t)\eta_a(s) \rangle =\frac{D_a}{\tau}\exp{\left( -|t-s|/\tau  \right)} \,.
\end{equation*}

 We point out that the relevance of the activity is determined by the ratio between the persistence time of the activity, $\tau$, and the time related to the external potential\cite{caprini2018activity}, $t_\Psi=\gamma/|\Psi'|'$. Indeed, when $\tau\ll t_\Psi$, we can approximate the noise as Brownian motion, $\eta_a\approx\sqrt{2D_a}\xi$, being $\xi$ a white noise, this corresponds to an adiabatic elimination of the faster degree of freedom. In particular, we obtain exactly this regime when $\tau=0$.

In the AOUP model,  $T^{-1}$ is a differential operator. As shown in Appendix \ref{appendixa}
it has the form:
\begin{equation}
\label{eq:Tmenus1Tb0}
T^{-1}(t)=\delta(t)G^{-1}(t)=\frac{\delta(t)}{2 D_a } \left( 1 - \tau^2\frac{d^2}{dt^2}\right)\,.
\end{equation}
For $\tau=0$, $T^{-1}$ reduces to a multiplicative operator, in agreement with the case of a Brownian suspension of particles with diffusion coefficient, $D_a$.
By using Eq.\eqref{eq:generaloverdamped}, identifying $\boldsymbol{\omega}$ with the particle position, $x$, and the vector force as $\Psi'=-F$, the entropy production rate of the system reads:
\begin{eqnarray}
\label{eq:entropyprodDt0}
\sigma(t)&&=-\frac{1}{\gamma}\biggl[\dot{x}(t) \left(T^{-1} \ast \Psi'\right)(t)  + \Psi'(t) \left(T^{-1} \ast \dot{x}\right)(t) \biggr]\nonumber\\
&&=-\frac{1}{2D_a\gamma} \biggl[\dot{x}(t) \int ds \delta(t-s)\left( 1- \tau^2 \frac{d^2}{ds^2}  \right) \Psi'(s)  + \Psi'(t) \int ds \delta(t-s)\left( 1- \tau^2 \frac{d^2}{ds^2}  \right) \dot{x}(s) \biggr]\nonumber\\
&&=-\frac{1}{D_a\gamma} \frac{d}{dt}\Psi + \frac{\tau^2}{2D_a\gamma}\biggl[ \dot{x}(t)\int ds \,\,\delta(t-s)\, \frac{d^2}{ds^2}\Psi'(s) + \Psi'(t)\int ds\,\, \delta(t-s)\,   \frac{d^2}{ds^2} \dot{x}(s) \biggr]\nonumber\\
&&=-\frac{1}{D_a\gamma} \frac{d}{dt}\Psi + \frac{\tau^2}{2D_a\gamma}\biggl[ \dot{x}^3(t)\Psi'''(t) + \dot{x}(t)\ddot{x}(t)\Psi''(t) +  \Psi'(t)    \frac{d}{dt}\ddot{x}(t) \biggr]\nonumber\\
&&=-\frac{1}{D_a\gamma} \frac{d}{dt}\Psi + \frac{\tau^2}{2D_a\gamma}\biggl[ \dot{x}^3(t)\Psi'''(t) + \frac{d}{dt}(\ddot{x}(t)\Psi'(t)) \biggr] \,.
\end{eqnarray} 
Formula \eqref{eq:entropyprodDt0} explicitly solves the recent dispute concerning the entropy production in the AOUP model, since 
it has been derived without arbitrary prescription regarding the parity of the self-propulsion force at variance with Mandal et al.~\cite{mandal2018}.
Their approach considers the introduction of the auxiliary variable $v\equiv\dot{x}$ and then, in analogy with the procedure adopted in the case of a magnetic force, arbitrarily changes the time-reversed path generator. The problem of this approach is discussed in~\cite{caprini2018comment}.

Going back to Eq.\eqref{eq:entropyprodDt0} we obtain the following formula for the entropy production rate:
\begin{equation}
\label{eq:entropy_vaccum}
\Sigma_{\mathcal{T}}=\int^{\mathcal{T}} ds \sigma(s) =\frac{\tau^2}{2D_a\gamma} \langle\dot{x}^3\Psi'''  \rangle 	\,\, {\mathcal{T}} + b.t.
\end{equation}
where the symbol $b.t.$ stands for boundary terms. The entropy production of the model vanishes when $\Psi'''(x)=0$, 
for instance in the case of a harmonically confined  active particle. Although this result does not apply to real bacteria, which are always out of equilibrium even in absence of any confining mechanism, it applies to this particular model~\cite{caprini2018comment}. 
 Instead, for a general trapping mechanism, Eq.\eqref{eq:entropy_vaccum} predicts a positive entropy production whose value depends on the third derivative of the potential. The $\tau$ dependence agrees with our intuition: 
when $\tau$ grows the distance from equilibrium grows too, corresponding to a larger entropy production.

Formula~\eqref{eq:entropy_vaccum} coincides with the one obtained by Fodor et al.~\cite{fodor2016far}, apart from irrelevant boundary terms, and shows that their strategy of not fixing the parity of the self-propulsion is the only possible one. 
 Following a different strategy, based on the
 method introduced by Seifert \cite{seifert2012stochastic} that identifies the entropy production of the medium directly from the manipulation of the
 Fokker-Planck equation,  Marconi et al found a result \cite{marconi2017heat} for the average entropy production rate which coincides with result~\eqref{eq:entropy_vaccum} except for irrelevant boundary terms as shown in the Appendix~\ref{appendixb}. 
Finally, using Eq.\eqref{eq:entropyprodrate} we have the dissipation of the AOUP model, which resembles the dissipation of a particle in a viscoelastic bath:
\begin{eqnarray}
I \, &&= -\dot{x} G^{-1} \Psi' - \Psi' G^{-1}\dot{x}  = -\frac{1}{2 D_a \gamma^2 } \left[\dot{x} \left( 1 - \tau^2\frac{d^2}{dt^2}\right) \Psi' + \Psi' \left( 1 - \tau^2\frac{d^2}{dt^2}\right)\dot{x}\right]\nonumber\\
&&=-\frac{\dot{x}\Psi'}{D_a \gamma^2 }  + \frac{\tau^2}{2D_a \gamma^2 }\left[ \dot{x}\frac{d^2}{dt^2}\Psi' + \Psi'\frac{d^2}{dt^2} \dot{x}   \right] \, .
\end{eqnarray}
Such a formula contains an extra term with respect to Fodor et al.~\cite{fodor2016far}, which stems from the second term of Eq.~\eqref{eq:entropyprodrate}.

\subsection{Example III: Self-propelled particles in a suspension at fixed temperature}
In a typical experimental setup, the self-propelled particles swim through a solvent, which 
is usually described as a medium at thermodynamic equilibrium. In order to model the solvent as a thermal bath, we assume that the microswimmers cannot change the equilibrium properties of the environment. 
In this picture, the self-propelled object is in an equilibrium-like regime with respect to the thermal reservoir but is intrinsically far from equilibrium as far as the source of active noise is involved.

Taking explicitly into account the role of the thermal bath could lead to non-trivial phenomena \cite{caprini2018active}, since formally the particle can be imagined in contact with two reservoirs, the active and the thermal one. For this reason, we consider the dynamics of a self-propelled particles, taking explicitly into account the role of the thermal bath, due to the solvent in which the active particle is immersed. We must consider the dynamics described by Eq.~\eqref{eq:generaloverdamped}
and replace $\eta\rightarrow \eta_a+\eta_t$, where $\eta_t$ is a white noise, such that $\langle \eta_t \rangle=0$ and $\langle \eta_t(t)  \eta_t(s) \rangle=2\,(T_b/\gamma)\,\delta(t-s)$, and  $\eta_a$ is the colored noise already introduced in subsection~\ref{notranslational}. 
Coherently with the previous notation we obtain that 
\begin{equation*}
\nu(t-s)=2\,\frac{T_b}{\gamma} \delta(t-s)+ \frac{D_a }{\tau} \exp{\left(-|t-s|/\tau\right)} \, ,
\end{equation*}
so that as shown in the Appendix~\ref{appendixa} we obtain the following expression for $T^{-1}$:
\begin{equation}
T^{-1}(t)=\frac{\gamma}{2T_b}\delta(t)  + K^{-1}(t)
\label{tm1}
\end{equation}
where 
\begin{equation}
 K^{-1}(t)=- \frac{ D_a \gamma^2}{2 T_b^2 } \left(\frac{1}{1+\frac{D_a\gamma}{T_b}}\right) \left[\frac{1}{\tau} \sqrt{1+\frac{D_a\gamma}{T_b}}\exp{\left( -\frac{|t|}{\tau} \sqrt{1+\frac{D_a\gamma}{T_b}}\right)}\right] \, .
 \label{tm2}
\end{equation}
 The first term in Eq.~\eqref{tm1} is only determined by the thermal noise,  while the second addend 
 depends both on the thermal and the active noises, 
Moreover, $T^{-1}$  is not proportional to a time $\delta$-Dirac function
 but is non-local in time as clearly shown by Eq.~\eqref{tm2}, displaying a time-exponential decay with typical time $\tau /\sqrt{1+D_a\gamma/T_b}$.
The above result has been independently derived in \cite{dabelow2018irreversibility}.
By applying the formula \eqref{eq:entropyprodrate}, the entropy production rate reads:
\begin{eqnarray}
\label{eq:entropyDt}
\sigma(t)&&=-\frac{1}{T_b} \dot{x}(t)\Psi'(x)(t) - \frac{\Psi'(t)}{\gamma}\int K^{-1}(t-s) \dot{x}(s) ds - \dot{x}(t)\int K^{-1}(t-s) 	\frac{\Psi'(s)}{\gamma} ds \nonumber\\
&&=-\frac{\Psi'(t)}{\gamma}\int K^{-1}(t-s) \dot{x}(s) ds - \dot{x}(t)\int K^{-1}(t-s) \frac{\Psi'(s)}{\gamma} ds  + b.t.
\end{eqnarray}
By considering the equilibrium limit, $\tau\rightarrow 0$, we expect that the entropy production is simply given by  a boundary term. 
Indeed, for $\tau\rightarrow0$ the expression contained in the square brackets of Eq.~\eqref{tm2} 
is proportional to a $\delta$-Dirac function. After some simple algebra, we get:
\begin{equation}
\label{eq:tinv}
\lim_{\tau\rightarrow 0}T^{-1}(t) = \frac{\delta(t)}{\left(D_a+ T_b/\gamma  \right)} \, ,
\end{equation}
which shows that the denominator is proportional to the sum of the temperature of the solvent and of the effective active temperature $D_a \gamma$, as we expect. Applying the form  \eqref{eq:tinv} of the operator $T^{-1}(t)$ in the entropy formula~\eqref{eq:entropyprodrate} we simply obtain a boundary term in analogy with the Brownian result
\eqref{sigmabrown}.
It is interesting to consider formula \eqref{eq:entropyDt} also in the limit $D_a\gamma/T_b \ll 1$. Since the amplitude of $K^{-1}(t)$ is order $O(D_a\gamma/T_b)$ the active entropy production rate decreases with this ratio,
as one should expect.

Finally, we discuss the singular limit $T_b\rightarrow 0$, where the source of thermal noise approaches to zero.  Neglecting the boundary terms we can show that:
\begin{equation}
\label{eq:entropyprod_limitTb}
\lim_{T_b \rightarrow0} \Sigma_{\mathcal{T}}(T_b \neq 0) =  \frac{\tau^2}{2D_a\gamma} \langle\dot{x}^3\Psi'''  \rangle 	\,\, {\mathcal{T}} ,
\end{equation}
which coincides with the result of  Eq.\eqref{eq:entropy_vaccum} obtained directly from the dynamics without thermal noise.
The explicit calculations are reported in Appendix~\ref{appendixc}.

 Due to the exponential shape of the memory kernel, we can argue that the main contribution of the entropy production stems from the short time region, where $t\sim s$. We thus expand the exponential kernel in powers of $|t-s|/\tau_R\ll1$, being $\tau_R$ the correlation time associated to $T^{-1}$, selecting the first order as the leading contribution. In the same way, the integral gives contribution only within $[t-\tau_R, t]$.
This idea can also be spelled by  saying that for each time history the largest contribution to the entropy production  occurs when the variables are strongly correlated.
In this way, we obtain:
\begin{eqnarray}
\label{eq:approx_entropy_Tb}
\sigma(t)  && = -2\frac{\Psi'(t)}{\gamma}\int^t_{t_0} K^{-1}(t-s) \dot{x}(s) ds -2\dot{x}(t)\int^t_{t_0} K^{-1}(t-s) \frac{\Psi'(s)}{\gamma} ds \nonumber\\
&&\approx 2\frac{A}{\gamma \tau_R}\left[\Psi'(t) \int^{ t}_{t-\tau_R} \dot{x}(s) ds + \dot{x}(t) \int^{t}_{t-\tau_R} \Psi'(s)ds \right] \nonumber\\
&&\approx 2\frac{A}{\gamma}\Bigl\{ \Psi'(t) \left[ \frac{x(t) - x(t-\tau_R)}{\tau_R}\right] -  \dot{x}(t)  \left[\Psi'(t) - \Psi'(t-\tau_R)\right]  \Bigr\} \, ,
\end{eqnarray}
being $A$ a constant and $\tau_R$ the correlation time of $T^{-1}$, given by:
\begin{equation}
A=\frac{ D_a \gamma^2}{2 T_b^2 } \left(\frac{1}{1+\frac{D_a\gamma}{T_b}}\right)\,, \qquad \tau_R=\tau \left(\sqrt{1+\frac{D_a\gamma}{T_b}}\right)^{-1} \,.
\end{equation}
In the second step of Eq.\eqref{eq:approx_entropy_Tb} we have evaluated the second integrals in the simplest way as possible. Formula \eqref{eq:approx_entropy_Tb} is consistent with what we expect in the limit $\tau\rightarrow0$. Indeed, when $\tau\rightarrow 0$ then $\tau_R\rightarrow 0$ and the first addend of Eq.\eqref{eq:approx_entropy_Tb} reduces to $\propto \Psi' \dot{x}$, i.e. to a boundary term. Moreover, the second term of Eq.\eqref{eq:approx_entropy_Tb} vanishes in this limit, meaning that the entropy production is zero consistently with the previous result.

Passing to the two time state variables probability, and using that the position $x(t-\tau_R)$ and $x(t)$ are in good approximation uncorrelated, i.e. $p(x(t-\tau_R), x(t)) \approx p(x(t-\tau_R)) p(x(t))$, we can write the average entropy production rate in a suitable way:
\begin{equation}
\label{eq:approx_averageentropyprodTb0}
\langle \sigma\rangle \approx 2\frac{A}{\tau_R\gamma} \left[\langle \Psi'(x) x \rangle - \langle x \rangle \langle \Psi'(x)\rangle  + \tau_R\langle \dot{x}\rangle \langle \Psi'(x) \rangle\right] \,.
\end{equation}
Noting that the third addend of Eq.\eqref{eq:approx_entropy_Tb} is just a boundary term.

We point out that at variance with the result of subsection \ref{notranslational}, the entropy production does not vanish, even in the presence of a harmonic potential, consistently with the fact that the system is now exchanging energy with two
different baths. Moreover, in this case, the average entropy production at the leading order is proportional to the $x$-variance of the process, being $\langle \dot{x} \rangle=0$, as explicitly evaluated in \cite{caprini2018active}. 

We remark that in the absence of external potentials the above formula independently of the form of Eq.~\eqref{eq:entropyprodrate}  displays a zero entropy production. As discussed in the previous section, the cause of this failure to describe the real behavior of active matter system is the poor modeling and  not related to the path integral definition of the entropy production, which is unique once the dynamics is specified.

\section{Conclusions}\label{sec:conclusion}
In this work, we  studied the entropy production of a target particle, under the action of one or more sources of noise - with and without  memory - following a path integral approach generalized to non-Markovian noises.
In particular, we discussed the case of an active particle moving  in absence of thermal noise. Several authors have presented different results concerning the entropy production rate,
a situation which provoked an interesting debate. Our approach does not involve arbitrary assumptions regarding the parity under TRT of the self-propulsion force,  which is neither even neither odd. Moreover, we confirm the result found with independent methods in \cite{fodor2016far} and \cite{marconi2017heat} apart from boundary terms. We employ the method also in a more general case, which takes into account the presence of the thermal bath. This leads to the calculation of the entropy production in  more realistic systems.

 Path-integral techniques have been recently used to compute the response due to a small perturbation \cite{szamel2017evaluating}, for instance a small shear flow \cite{asheichyk2018response}. Our non-markovian techniques could provide an interesting way to compute such observables.
 
\appendix
\section{Computation of $\boldsymbol{T}^{-1}$}
\label{appendixa}
The computation of the operator $\boldsymbol{T}^{-1}$, i.e. the inverse of the correlator $\boldsymbol{\nu}$, follows directly from the definition of the inverse operator:
\begin{equation}
\int  \boldsymbol{T}^{-1}(t-t') \boldsymbol{\nu}(t'-s) dt' = {\boldsymbol{\delta}}(t-s) \,.
\end{equation}
Thus, $\boldsymbol{T}^{-1}$ can be easily evaluated in the Fourier space. Introducing $\mathcal{FT}$ as the Fourier trasform operator and $\mathcal{FT}^{-1}$ as the inverse-Fourier trasform operator  and denoting by a tilde the Fourier transform of a given function, we have by definition that:
\begin{equation}
\label{eq:Tmenus1general}
\boldsymbol{T}^{-1}(t) = \mathcal{FT}^{-1}(\tilde{\boldsymbol{T}}^{-1}) =\int \frac{d\omega}{2\pi} \tilde{\boldsymbol{T}}^{-1}(\omega) e^{-i\omega t} \,.
\end{equation}
The convolution in $\mathcal{FT}$-space reads:
\begin{equation}
\tilde{\boldsymbol{T}}^{-1}(\omega)=\frac{1}{\tilde{\boldsymbol{\nu}}(\omega)}, \qquad \tilde{\boldsymbol{\nu}}(\omega)=\int dt e^{i\omega t}\boldsymbol{\nu}(t) \,.
\end{equation}
Applying such a prescription we can evaluate the operator $\boldsymbol{T}^{-1}$ in all the cases discussed in the previous Sections. For the sake of simplicity we consider the scalar case, without loss of generality:
\begin{enumerate}
\item Brownian particle. The dynamics can be obtained by the general form \eqref{eq:generaloverdamped}, by replacing $\eta$ with $\sqrt{2T_b/\gamma}\,\eta$.
Applying Eqs.\eqref{eq:Tmenus1general} we get:
\begin{equation}
\tilde{\nu}(\omega) = 2 T_b, \qquad T^{-1} = \frac{\delta(t)}{2T_b} \,.
\end{equation}

\item Active particle without thermal noise. In this case the dynamics is done by assuming that in Eq.\eqref{eq:generaloverdamped} the noise $\eta$ is replaced with $\rightarrow \eta_a$, an object with  zero average and two time correlation $\nu(t-s) = \gamma^2 (D_a/\tau) e^{-|t-s|/\tau}$. In this case, Eq.\eqref{eq:Tmenus1general} gives us the following results:
\begin{equation}
\label{eq:AentropyDt0}
\tilde{\nu}(\omega)=\frac{2D_a}{1+\tau^2\omega^2}, \qquad T^{-1}(t)=\frac{\delta(t)}{2 D_a  } \left( 1 - \tau^2\frac{d^2}{dt^2}\right).
\end{equation}
We note that $T^{-1}$ is a differential operator, which is even in its argument.

\item Active particle in contact with a thermal bath. In this case $\eta$ in Eq.\eqref{eq:generaloverdamped} has to be replaced by the sum of the previous terms, $\eta \rightarrow \sqrt{2T_b/\gamma}\,\eta +  \eta_a$. In this way the correlation is simply given by $\nu(t-s) = 2\,(T_b/\gamma) \delta(t-s)+ (D_a/\tau) e^{-|t-s|/\tau}$ and so applying Eq.\eqref{eq:Tmenus1general} we get:
\begin{eqnarray}
\label{eq:AentropyDt}
&&
\tilde{\nu}(\omega) = \frac{2D_a }{1+\tau^2\omega^2} + \frac{2T_b}{\gamma}, 
\\&& T^{-1}(t)=\frac{\gamma}{2T_b}\delta(t) - \frac{ D_a \gamma^2}{2 T_b^2 } \left(\frac{1}{1+D_a\gamma/T_b}\right) \left[\frac{1}{\tau} \sqrt{1+\frac{D_a\gamma}{T_b}}\exp{\left( -\frac{|t|}{\tau} \sqrt{1+\frac{D_a\gamma}{T_b}}\right)}\right]. \nonumber\\
\end{eqnarray}
The presence of two baths leads to an operator which is not proportional to a $\delta$ Dirac function. 
\end{enumerate}
As a last remark, all the previous calculations and definitions can be easily generalized to the interacting case in more dimensions without complications of any kinds. 

\section{ Equivalence between the Fokker-Planck appproach and the path integral approach in the computation of the entropy production}
\label{appendixb}
Following Marconi et. al.~\cite{marconi2017heat}, we can compute the entropy production rate of the medium associated to the AOUP dynamics in absence of thermal noise, by using a Fokker-Planck approach. 

By specifying the evolution of $\eta^a$ as an Ornstein-Uhlenbeck process, we can easily map the active overdamped dynamics into  the underdamped dynamics of a fictitious Brownian particle, where the activity is mapped into a space dependent Stokes force~\cite{marconi2016velocity}:
\begin{equation}
\label{eq:B1}
\ddot{x}=-\dot{x} \frac{\Gamma(x)}{\tau} - \frac{\Psi'(x)}{\tau} + \frac{\sqrt{2D_a}}{\tau}\xi\,, \qquad \Gamma=1+\tau \Psi''(x)\,,
\end{equation}
where $\xi$ is the white noise, such that $\langle \eta \rangle=0$ and $\langle \xi(t)\xi(t') \rangle=\delta(t-t')$.

Following~\cite{marconi2017heat}, we find an expression for the entropy production of the medium (by setting $\gamma=1$ for simplicity):
\begin{equation}
\label{eq:AEntrProd}
\langle \sigma(t)\rangle = \int dx d\dot{x} \frac{\tau}{D_a}\Gamma(x)\left[\frac{\Gamma(x)}{\tau}\dot{x}^2p(x,\dot{x}) -\frac{D_a}{\tau^2}p(x,\dot{x})  \right],\qquad \Gamma=1+\tau U''(x) \,.
\end{equation}

In order to proceed further, we develop a formal relation, by comparing different formulations of the linear response theory.
In particular, in ref.~\cite{marconi2008fluctuation} the linear response of an observable $O$, which depends on the dynamical variables, due to a small force perturbation, $h$, reads:
\begin{equation}
\label{eq:VulpResp}
\mathcal{R}_{Ov}(t-s)=-\langle O(t) \frac{d}{dv} \log{p(x,\dot{x})}(s) \rangle \,,
\end{equation}
being $p$ the probability in the space $(x,v)$, which is actually unknown. Recently, other relations have been developed \cite{speck2006restoring,speck2010driven}, for instance:
\begin{equation}
\label{eq:OtherResp}
\mathcal{R}_{Ov}(t-s) = \frac{2\tau}{\sqrt{2D_a}}\langle O(t)\xi (s) \rangle \,,
\end{equation}
adapting the coefficient to the specific case we are considering. Then, identifying the rhs of Eqs.\eqref{eq:VulpResp} and \eqref{eq:OtherResp} and since these relations are valid for the generic observable $O(t)$, we can find an integral equation which connect $p(x,\dot{x})$ and $\xi$:
\begin{equation}
\label{eq:B5}
\int dx d{\dot{x}}\frac{\sqrt{2}\tau}{\sqrt{D_a}}\xi\,\, p(x,\dot{x}) \,O(x, \dot{x}) = - \int dx d{\dot{x}} \frac{\partial}{\partial\dot{x}}p(x,\dot{x}) \,O(x, \dot{x}) \,.
\end{equation}
By taking intou account this relation, Eq.\eqref{eq:AEntrProd} reads:
\begin{eqnarray}
\label{eq:B6}
\langle \sigma(t)\rangle &&= \int dx d\dot{x} \frac{\tau}{D_a}\Gamma(x)\left[\frac{\Gamma(x)}{\tau}\dot{x}^2 -\frac{D_a}{\tau^2}  \right]p(x,\dot{x}) \nonumber\\
&&= \int dx d\dot{x} \frac{\tau}{D_a}\Gamma(x)\dot{x}\left[\frac{\Gamma(x)}{\tau}\dot{x}\,\,p(x,\dot{x}) +\frac{D_a}{\tau^2}\frac{\partial}{\partial \dot{x}}p(x,\dot{x})  \right] \nonumber \\
&&= \int dx d\dot{x} \frac{\tau}{D_a}\Gamma(x)\dot{x}\left[\frac{\Gamma(x)}{\tau}\dot{x} -\frac{\sqrt{2D_a}}{\tau}\,\,\xi\,\,  \right]\,p(x,\dot{x})=- \frac{1}{\mathcal{T}}\int^{\mathcal{T}} dt \frac{\tau}{D_a}\Gamma(x)\dot{x}\left[\ddot{x} + \frac{\Psi'}{\tau}  \right]\nonumber \\
&&=- \frac{1}{\mathcal{T}}\int^{\mathcal{T}} dt \frac{\tau}{D_a}\left[1+\tau \Psi''(x)\right]\dot{x}\left[\ddot{x} + \frac{\Psi'}{\tau}  \right] \nonumber\\
&&=- \frac{1}{\mathcal{T}}\int^{\mathcal{T}} dt \frac{\tau}{D_a}\left[ \frac{d}{dt}\frac{\dot{x}^2}{2}  +\tau\,\dot{x}\,\ddot{x} \Psi''(x) + \frac{d}{dt}\frac{\Psi(x)}{\tau} + \frac{d}{dt}\frac{\Psi'^2}{2}   \right] \nonumber\\
&&= - \frac{1}{\mathcal{T}}\int^{\mathcal{T}} dt \frac{\tau^2}{2D_a}\dot{x}^3 \Psi'''(x) + b.t. 
\end{eqnarray}
where $b.t.$ stems for boundary terms. We remark that in the second equality of Eq.\eqref{eq:B6} we have integrated by parts,  in the third we have used Eq.\eqref{eq:B5} and in the fourth we have replaced the square brackets with Eq.\eqref{eq:B1} and we have switched to the time integral assuming the ergodicy of such a system, ${\cal T} \,\int dx d\dot{x} p(x,\dot{x}) =\int^{\cal T} dt $. 
 Last equalities are just the result of an integration by parts.
 As we claim, Eq.\eqref{eq:B6} coincides with the one obtained in \cite{fodor2016far},
 confirming the consistency of the two approaches.

\section{Entropy production in the limit of zero thermal noise}
\label{appendixc}
In order to demonstrate Eq.~\eqref{eq:entropyprod_limitTb} it is sufficient to prove that:
\begin{equation*}
\lim_{T_b \rightarrow0} K^{-1}(t)\approx -\frac{\tau^2}{2 D_a} \frac{d^2}{dt^2}  \delta(t)  \equiv -\frac{\tau^2}{2 D_a} \delta(t) \frac{d^2}{dt^2} \,,
\end{equation*}
where the symbol $\approx$ means that we neglect additive contributions only representing boundary terms in the entropy production, in particular addends which are proportional to $ \delta(t)$.
Using the identity:
\begin{equation*}
\frac{d^2}{dt^2} \left[\frac{1}{\tau_R} \exp{\left(- \frac{ t}{\tau_R}   \right)}\right] = \frac{1}{\tau_R^2}  \left[\frac{1}{\tau_R}\exp{\left(-  \frac{ t}{\tau_R}  \right)}\right],
\end{equation*}
$K^{-1}(t)$ given by \eqref{tm2} can be conveniently rewritten as:
\begin{equation*}
K^{-1}(t) =- \frac{ D_a \gamma^2}{2 T_b^2 } \left(\frac{1}{1+\frac{D_a\gamma}{T_b}}\right)^2\frac{d^2}{dt^2}  \left[\frac{1}{\tau_R} \exp{\left(- \frac{ t}{\tau_R}   \right)}\right] \,,
\qquad \tau_R   =\frac{\tau}{ \sqrt{1+\frac{D_a\gamma}{T_b}}} \,.
\end{equation*}
As $T_b \rightarrow 0$ then $\tau_R\rightarrow 0$ and the expression inside the square brackets tends to a $\delta$-Dirac function, while the prefactor reduces to $- \tau^2/2D_a$.



\section*{References}

\providecommand{\newblock}{}

\end{document}